# Folding an e-graph in pure egglog

Zeng Ren
EPFL

Maddy Bowers
MIT

## Abstract

Folding over a recursive data structure (also known as "catamorphism") is an essential operation in functional programming[1]. It allows one to declaratively define a recursive computation by focusing on how solutions of subproblems are combined into the solution of the parent problem. Folding in the context of e-graphs can have several benefits. Not only do we get memoization for free, but also computation is drastically reduced since the subproblems are defined over e-classes as opposed to concrete sub-terms. This short report presents our initial attempt at defining and implementing catamorphism, a general notion of a fold, in pure **egglog**. We discuss a reusable implementation trick to stage the inference rules so that there is no re-propagation of updated values.

## 1 Introduction

Many analyses on e-graph over a least fixed-point type Fix $t$ are defined as a mutual recursion between e-node values and e-class values.[1] The e-node computation $f : t\ a \to a$ specifies how values of child e-class is combined in to the value of the e-node, while the e-class computation $g$ : Multiset $a \to a$ specifies how a collection of e-node values is combined into the values of the e-class consisting of these e-nodes.

This abstract description fits many useful bottom-up analyses on e-graphs. For example, in Babble's e-class anti-unification algorithm[2], the $g$ part corresponds to the `Dominant` function, which filters a collection of patterns based on a partial order, while $f$ part follows from traditional term-wise anti-unification. Another example is in Babble's cost-set analysis algorithm, where the $g$ part corresponds to the partial order reduction followed by top-k pruning, while the $f$ part carries the core logic of reasoning about the set of candidate library and its use cost. These two algorithms were originally implemented using **egg**[3] (and before egglog was published) while the bulk of the algorithms logic is in rust. Can we implement such algorithms declaratively in pure **egglog**[4]?

Current **egglog** is built-on the assumption that the e-class analysis value is defined as the aggregated e-node values under a custom binary merge function. But there are some analysis on e-classes, besides the two examples from Babble that depends on all the its e-node analysis values.

Here we sketch out an implementation of such general catamorphisms on e-graphs in pure **egglog**.

## 2 Folding an e-graph

We need to split e-class analysis $F$ into two functions: $F_{\text{Final}}$ denotes the finalized analysis of an e-class (the result of after the merge function $g$), and $F_{\text{on-going}}$ denotes the building up multi-set of e-node analysis of a e-class.

For the e-node values function $f$, it can be translated to a collection of **egglog** rules (see Figure 1), whose premises establish binding to the values for sub-term (e-class), and whose action inserts a e-node analysis value to its corresponding e-class aggregated value set.[2] There will be a rule for each constructor of the datatype, mirroring the different cases for of the function $f$ to pattern match on. Let's call this rule set "inductive-rules".

$$\frac{\begin{array}{c} e = C_{\text{Fix}}\ x_1\ \ldots\ x_n \\ y_1 = F_{\text{Final}}\ x_1 \\ \vdots \\ y_n = F_{\text{Final}}\ x_n \end{array}}{F_{\text{on-going}}\ e \overset{\cup}{=} \{f\ (C_{\text{Base}}\ y_1\ \ldots\ y_n)\}}\ \textsc{Inductive}$$

Figure 1: A first step translation of e-node value computation (parametrized by an algebra $f$) to an **egglog** rule. $C_{\text{Fix}}$ as a constructor of the fixed-point type, whose base functor has the corresponding constructor $C_{\text{Base}}$

For the e-class value computation, we also have a simple translation (see Figure 2). Let's call this rule set "resolve-rules" as it resolves the functional dependency for **egglog** functions.

[1] This is natural since each e-node points to e-classes, which further contains multiple e-nodes.

[2] It might be tempting to omit the first premise and just inline $\alpha$ in the concluding action. This will have the unintended effect of introducing additional terms (of the domain type) to the e-graph indefinitely. The first premise is there to say that we only want to compute the analysis on terms that are present in the e-graph.

$$\frac{e = C_{\text{Fix}}\ x_1\ \dots\ x_n}{F_{\text{Final}}\ e = g\ \left(F_{\text{on-going}}\ e\right)}\ \text{RESOLVE}$$

Figure 2: A first step translation of e-class value computation (parametrized by a merge function $g$) to an **egglog** rule.

Note that this resolve rule (as well as $F_{\text{on-going}}$) is not required when our function g can be incrementalized by iterating a binary merge function, which is the natively supported behavior in **egglog**.[3]

To compute all the $F_{\text{Final}}$ for all the e-classes of a given type, it can be tempting to schedule the rules as

```
(saturate
  (run inductive-rules)
  (saturate eval-f-rules)
     ^ if f is defined as a constructor
  (run resolve-rules)
  (saturate eval-g-rules)
     ^ if g is defined as a constructor
)
```

Listing 1: A tempting schedule that is error prone

However this comes with some race condition issues. The resolve rules should only trigger when all the e-nodes values are computed, otherwise we would propagate a wrong value. To fixed this, we need a way to express an e-class have all of its e-nodes' values ready.

## 3 Rule scheduling

We want the order of computation to respect the data dependency of the mutual recursion, in particular: the resolve rule to be triggered on an e-class $e$ only after all of its e-nodes values are computed.

We can try to fix this by introducing a relation `AllAnalyzed` to witness an e-class has all of its e-node analyzed. We can use a multi-set to represent all the e-nodes analysis of an e-class (as multiplicity matters and order does not matter for our merge function $g$). Suppose our domain datatype is $A$ and analysis has type $B$ then the relation can be declared as

```
(sort MultiSetB (multi-set B))
(relation AllAnalyzed (A MultiSetB))
```

This new relation should be supplied as an additional premise to the `resolve-ruleset` as

$$\frac{\begin{array}{c} e = C_{\text{Fix}}\ x_1\ \dots\ x_n \\ \text{AllAnalyzed}\ e\ S \end{array}}{F_{\text{Final}}\ e = g\ S}\ \text{RESOLVE}'$$

Figure 3: We add a new entry to $F_{\text{Final}}$ on class $e$ when all of its e-nodes has been analyzed

### 3.1 Checking the readiness of e-node analysis

How can one conclude the `AllAnalyzed` relation? One way is to keep track of the processed e-nodes (that has computed e-node analysis) with respect to the total amount the e-nodes in the e-class. This can be done in pure egglog by introducing two functions.

```
(function TotalNodes (A) i64 :merge (+ new old))
(function NodesDone (A) i64 :merge (+ new old))
```

The `TotalNodes` should be precomputed for e-classes of type $A$ by the inference rule in Figure 4.

$$\frac{e = C_{\text{Fix}}\ x_1\ \dots\ x_n}{\text{TotalNodes}\ e \stackrel{+}{=} 1}\ \text{CountNodes}$$

Figure 4: An e-class's number of e-nodes (with $C_{\text{Fix}}$ constructor) can be counted by triggering a counter increment on each matching term. This is valid as the rule will be matched for each e-node with the $C_{\text{Fix}}$ constructor, and all **egglog** functions are defined over e-classes.

On the other hand, the `NodesDone` function, which represents the progress of e-node analysis within a e-class, should be initialized at 0 and increment whenever $f_{\text{on-going}}$ is updated (see Figure 5).

$$\frac{\begin{array}{c} e = C_{\text{Fix}}\ x_1\ \dots\ x_n \\ y_1 = F_{\text{Final}}\ x_1 \\ \vdots \\ y_n = F_{\text{Final}}\ x_n \end{array}}{\begin{array}{l} F_{\text{on-going}}\ e \stackrel{\cup}{=} \{f\ (C_{\text{Base}}\ y_1\ \dots\ y_n)\} \\ \text{NodesDone}\ e \stackrel{+}{=} 1 \end{array}}\ \text{INDUCTIVE}'$$

Figure 5: The new version of the inductive rule update the progress of the e-node analysis while the e-code analysis is aggregating.

The original call of $F_{\text{on-going}}$ has now moved to a new rule which manages the `AllAnalyzed` relation (see Figure 6.

$$\frac{\text{TotalNodes}\ e = \text{NodesDone}\ e}{\text{AllAnalyzed}\ e\ \left(F_{\text{on-going}}\ e\right)}\ \text{ConcludeAnalyzed}$$

Figure 6: We can freeze the accumulating e-node analysis of a e-class $e$ when the number analyzed nodes matches the total required nodes in $e$.

### 3.2 The final picture

Putting these rule-sets together, we have `CountNodes` to pre-computes the number of required e-nodes for each e-classes,

---

[3] An **egglog** function has value over a semi-lattice, which can be specified by a custom binary merge operation $_ \times _ : a \to a \to a$ that satisfy certain laws. If there are multiple computed value for e-nodes for the e-class, the e-class value is defined as folding over the set of values using the custom merge function $f(c) = \prod_{n \in c} n$.

There are many interesting applications that breaks structure of iterating the binary merge function, for example, taking the average, maintaining a bounded priority, updating the frontier of a set of values under a partial order.

`Inductive-rules` to update the function $F_{\text{on-going}}$ and `NodesDone`, which provides preconditions for `analyzed-rules` to conclude the whether a e-classes' e-nodes has been `AllAnalized`, which will trigger the `Resolve-rules` to generate new entries for $F_{\text{Finalized}}$ and thus starting a new cycle. The full scheduler is shown in Listing 4.

```
(run CountNodes-rules) ;; count TotalNode for e-classes
(saturate
  (run inductive-rules) ;; apply the custom algebra f on
already computed analysis on subterm e-classes (F-final e),
and insert the e-node analysis to its e-class value set
(F_ongoing)
  (saturate eval-f-rules) ;; fully evaluate custom algebra f
  (run analyzed-rules) ;; deduce AllAnalyzed relation
  (run resolve-rules)  ;; apply the custom merge function g
and set its result to F_final
  (saturate eval-g-rules) ;; fully evaluate custom merge
function g
)
```

Listing 4: The full schedular for the generic fold

## 4 Conclusion

This preliminary result is part of an effort to re-implement Babble in pure **egglog**. We find that there is a common “pattern” in extending recursive computation on terms to one involving e-classes. During this process we distilled the core problem down to implementing a general fold on e-graph parametrized a user defined algebra $f$ and a merge function $g$.

Here we presents a sketch for implementing a general fold on e-graph in pure **egglog** to model mutually recursive e-node and e-class analysis. The core issue we are tackling is to make sure the e-node and e-class analysis are computed in the right order, which is controlled by declarative **egglog** rules. We approach it by first separating the algebra function into a **egglog** function and a relation: the former for the on-going accumulation of e-node analysis, and the later to mark an e-class that has all its e-node analyzed. We only resolve the e-class analysis (using the user-defined merge function) after its e-nodes analysis are ready. We use a simple counter to check whether all e-nodes (that requires an analysis) are analyzed. This might be a point for future improvement as it can be sensitive to how **egglog** is evaluated under the hood. The whole fold can be computed by saturating interleaved runs of e-node and e-class analysis.

The code in our sketch here is parametrized over the particular algebra and merge function for the fold, which makes it easy to adapt to different kind of analysis on e-graphs.